\documentclass[12pt]{iopart}
\usepackage{bbm}
\expandafter\let\csname equation*\endcsname\relax
\expandafter\let\csname endequation*\endcsname\relax 
\usepackage{amsmath}
\usepackage{amssymb}
\usepackage{graphicx}
\usepackage{placeins}
\usepackage{epstopdf} 
\usepackage{subfigure}
\usepackage[utf8]{inputenc}

\newcommand{\Hc}{\mathcal{H}}
\newcommand{\Hid}[0]{\Hc_{\mathrm{id}}}

\newcommand{\Havg}[0]{\overline{\Hc}}
\newcommand{\Havgo}[0]{\Havg^{(0)}}
\newcommand{\id}[0]{\mathbbm{1}}
\newcommand{\ket}[1]{\left| #1 \right \rangle \!}

\newcommand{\braket}[2]{\left\langle #1 \vert #2 \right\rangle \!}
\newcommand{\abs}[1]{\left| #1 \right|}

\begin{document}
\title[]{Selective dynamical decoupling for quantum state transfer}
\author{H Frydrych$^1$, A Hoskovec $^2$, I Jex$^2$ and G Alber$^1$}
\address{$^1$ Institut für Angewandte Physik, Technische Universität Darmstadt, D-64289 Darmstadt, Germany}
\address{$^2$ Department of Physics, FNSPE, Czech Technical University in Prague,
B\v{r}ehov\'a 7, 115 19 Praha 1, Star\'e M\v{e}sto, Czech Republic}
\eads{\mailto{holger.frydrych@physik.tu-darmstadt.de}, \mailto{hoskoant@fjfi.cvut.cz}}

\begin{abstract}
State transfer across discrete quantum networks is one of the elementary tasks of quantum information processing. Its aim is the faithful placement of information into a specific position in the network. However, all physical systems suffer from imperfections, which can severely limit the transfer fidelity. We present selective dynamical decoupling schemes which are capable of stabilizing imperfect quantum state transfer protocols on the model of a bent linear qubit chain. The efficiency of the schemes is tested and verified in numerical simulations on a number of realistic cases. The simulations demonstrate that these selective dynamical decoupling schemes are capable of suppressing unwanted errors in quantum state transfer protocols efficiently.
\end{abstract}


\submitto{\jpb}


\maketitle

\section{Introduction}
Transferring arbitrary quantum states is a task of central importance for quantum communication and also more generally for quantum information processing. 
As an arbitrary quantum state cannot be cloned perfectly \cite{wz82}, it is important to develop quantum state transfer protocols which are capable of transferring an arbitrary quantum state  within a quantum network from one position to any other.
Recently, quantum state transfer protocols have been developed 
independently by Bose \cite{bose}, Nikolopoulos \emph{et al.} \cite{nikolpetro} and Christandl \emph{et al.} \cite{christandl} for linear qubit chains. These protocols propose specific
Hamiltonians governing the dynamics of these chains
which implement a state transfer from one end of the chain to the other in a particular interaction time without any additional external control or ancillary quantum systems. A comprehensive introduction to the topic of quantum state transfer and current developments can be found in \cite{nikoljex, kay}.

Although the simplicity of these protocols is very appealing, they are susceptible to imperfections in the structure of the qubit chain. The effects of diagonal and off-diagonal disorder in the governing system Hamiltonian have been studied for spin chains in \cite{pnl2010,ybb2010,psb2014}.
Furthermore, while linear qubit chains with nearest neighbour interactions are convenient for exploring basic theoretical aspects of quantum state transfer,
experimental implementations typically involve more complicated and higher-dimensional scenarios. A particular arrangement, which can arise naturally in two- or three-dimensional qubit networks, is a qubit chain with a bend around a specific qubit. In such a case additional strong couplings
between qubits may arise close to the position of the bend so that simple one-dimensional models
with nearest-neighbour couplings no longer describe these situations adequately.
Such configurations have been studied recently in detail \cite{nikolhosk}. In particular, it has been demonstrated that the additional interactions arising from
qubits close to the position of the bend significantly affect quantum state transfer in a
detrimental way. Therefore, for practical implementations of quantum state transfer protocols it is important to develop techniques which suppress these detrimental
effects in an efficient way. 

A powerful technique to suppress unwanted interactions in quantum networks is selective dynamical decoupling, which has its origins in the context of nuclear magnetic resonance \cite{hahn, Carr, Meiboom, Haeberlen}. Viola {\it et al.} \cite{vkl99} later formalised these methods with particular emphasis on quantum information processing. Dynamical decoupling is based on the repeated application of appropriately chosen external control pulses to the physical system of interest. Unwanted interactions are suppressed by the resulting unitary transformations
which tend to average out large parts of these perturbations. This technique has been adopted in numerous experiments  in order to suppress unwanted dynamical influences on qubit systems \cite{mtabplb06, fsl05, budsib09, lgdvt11}. Although originally developed mainly for purposes of suppressing fidelity decay of open quantum systems originating from interactions with an environment \cite{vlk99}, in the meantime powerful systematic methods have also been developed for designing dynamical decoupling schemes for protecting an ideal system dynamics against unwanted perturbations \cite{fab14,cbk11}. Related concepts have been recently proposed to decouple a quantum state transfer from the effects of an environmental bath \cite{ebck11,zabk14}.

Motivated by the above mentioned recent developments in quantum state transfer, in this paper we present three selective dynamical decoupling schemes which are capable of suppressing efficiently unwanted qubit couplings of a bent qubit chain. These decoupling schemes require repeated applications of single-qubit Pauli pulses on the individual qubits, which on first sight contradicts the original idea of control-free state transfer protocols. Therefore, we expect our decoupling method to be suitable primarily for small to medium-sized qubit chains and particularly for state transfer within a quantum register where single-qubit gates are already available. In these scenarios, the schemes' exclusive dependency on Pauli pulses should make them particularly suitable for possible experimental applications. Of course, quantum state transfer can always be achieved by sequential swap operations between the neighbouring qubits, but the single-qubit level of control we assume is still much lower than the two-qubit swap gates would require. In a sense, the schemes presented here perform a state transfer with single qubit operations, which is otherwise impossible without modification of coupling strengths. The efficiency of our schemes is demonstrated by a number of numerical simulations.

The structure of the paper is as follows: In section \ref{sec:formalism} basic aspects of recently introduced quantum state transfer protocols are recapitulated which involve
linear and bent qubit chains. Section \ref{sec:methods} provides a brief summary of basic ideas of dynamical decoupling. In section \ref{sec:solution} three selective dynamical decoupling schemes are presented which are capable of suppressing effects of the unwanted qubit couplings in quantum state transfer along
a bent qubit chain. The first scheme is capable of protecting only the interaction part of the ideal Hamiltonian with the possibility that the qubit eigenenergies are rescaled in the process, whereas the more elaborate second scheme protects the ideal Hamiltonian completely. The simpler and more intuitive first decoupling scheme allows us to demonstrate
the basic ideas involved in a simple way. The second complete selective decoupling scheme presented
has been found by application of the previously developed systematic construction procedure \cite{fab14} and its functioning can be understood in a straightforward way on the basis of the simpler
and more intuitive first scheme. The final scheme is tailored to be particularly easy to implement experimentally and offers a certain robustness against diagonal disorder in the qubit chain.
Section \ref{sec:results} demonstrates the effectiveness of our proposed selective dynamical decoupling schemes with numerical simulations. 

\section{State transfer on qubit chains} \label{sec:formalism}
We associate a qubit with a quantum system with two orthogonal states $\ket{0}$ and $\ket{1}$ on a Hilbert space $\mathbb{C}^{2}$, on which any linear operator can be expressed as a linear combination of the unitary and Hermitian Pauli operators and the identity
\begin{align}
\sigma^{x} &=\begin{pmatrix}
0 & 1\\
1 & 0
\end{pmatrix}, & 
\sigma^{y} &=\begin{pmatrix}
0 & -i\\
i & 0
\end{pmatrix}, \notag \\
\sigma^{z} & =\begin{pmatrix}
1 & 0\\
0 & -1
\end{pmatrix}, & 
I & =\begin{pmatrix}
1 & 0\\
0 & 1
\end{pmatrix}.
\end{align}

A qubit network consists of $N$ distinguishable qubits spanning a Hilbert
space $\left(\mathbb{C}^{2}\right)^{\otimes N}$. Correspondingly, a qubit network is called
a linear chain if the qubits can be numbered from $1$ to $N$ such that any qubit $i$ only interacts with its direct neighbours $i\pm 1$. The qubits $1$ and $N$ are the ends of the chain and interact with only one neighbouring qubit.

In the following, we restrict ourselves to the study of an $XX$ type nearest-neighbour interaction on the qubit chain, which in the ideal case is given by a Hamiltonian (assuming $\hbar = 1$)
\begin{equation}
\Hid=\sum_{i=1}^N B_{i}\sigma_{i}^{z}-\sum_{i=1}^{N-1} J_i \left(\sigma_{i}^{x}\sigma_{i+1}^{x}+\sigma_{i}^{y}\sigma_{i+1}^{y}\right).
\label{eq:HHamiltonian}
\end{equation}
Here, $\sigma_{i}^k$ denotes $\sigma^k$ applied to the $i$-th qubit and the eigenenergies $B_i$
and coupling strengths $J_i$ are determined by the specific implementation of the qubit chain. It has been shown that for particular choices of the coupling strengths $J_i$ this Hamiltonian can transfer a single excitation from one end of the chain to the other one and thus can be used for purposes of perfect state transfer along the qubit chain. A particular choice for the coupling strengths $J_i$ has been proposed independently in \cite{nikolpetro} and \cite{christandl}, namely
\begin{equation}
J_i = \frac{\lambda}{2} \sqrt{i(N-i)} . \label{eq:couplings}
\end{equation}
If the qubit chain is prepared in the initial state \begin{equation}
\ket{\Psi(0)} = (a \ket{0} + b \ket{1}) \otimes \ket{0} \otimes \dots \otimes \ket{0}, \quad |a|^2 + |b|^2 = 1,
\end{equation} 
this particular choice of coupling strengths leads to the final state 
\begin{equation}
\ket{\Psi(T)} = \ket{0} \otimes \dots \otimes \ket{0} \otimes (a \ket{0} + e^{i\varphi} b \ket{1}),
\end{equation}
after a time $T = {\pi}/{\lambda}$. The phase $\varphi$ depends on the length of the chain $N$ and on the eigenenergies $B_i$ and should ideally be zero in order to accomplish perfect state transfer. Alternatively, the phase needs to be corrected by applying an appropriate phase gate at the end of the quantum state transfer. In the case where all $B_i = 0$, the phase is given by $e^{i\varphi} = (-i)^{N-1}$ (see \cite{pnl2010}). If the $B_i$ are non-zero, but uniform, $B_i = B$, they contribute an additional relative phase shift so that the final phase is given by 
\begin{equation}
e^{i\varphi} = (-i)^{N-1} e^{2 i B T}.
\end{equation}
This follows because in the case of uniform eigenenergies, the eigenenergy terms commute with the couplings in the Hamiltonian:
\begin{equation}
\left[ \sum_{i=1}^N B \sigma_i^z, \quad \sum_{i=1}^{N-1} J_i \left( \sigma_i^x \sigma_{i+1}^x + \sigma_i^y \sigma_{i+1}^y \right) \right] = 0.
\end{equation}
A choice of $B = (N-1)\lambda/4$ ensures that there is no phase shift. If the eigenenergies are different from each other, the state transfer is disturbed. The effects of this diagonal disorder were studied in \cite{pnl2010} and \cite{ybb2010}.

\begin{figure}
\includegraphics[width=7.5cm]{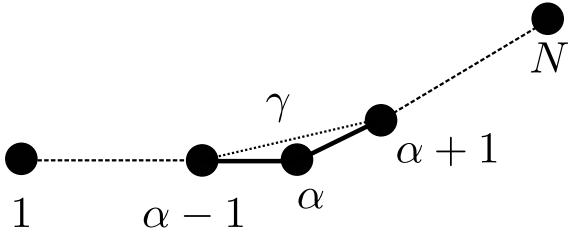}

\caption{Qubit network with an additional interaction at the bend}
\label{fig:bent}
\end{figure}
Let us now consider an additional interaction between qubits $\alpha-1$ and $\alpha+1$ as described by the Hamiltonian
\begin{equation}
\mathcal{H}=\Hid+\gamma\left(\sigma_{\alpha-1}^{x}\sigma_{\alpha+1}^{x}+\sigma_{\alpha-1}^{y}\sigma_{\alpha+1}^{y}\right),
\label{eq:hamiltonian}
\end{equation}
with $\gamma\in\mathbb{R}$ and $\alpha\in [2,N-1]$. This situation arises naturally if we consider a physical implementation of a qubit chain in which the coupling strengths between qubits are based on their physical distance. If there is a bend in the chain at qubit $\alpha$, it is conceivable that the coupling strength between the two neighbouring qubits at the bend becomes large enough so that it is no longer negligible (compare with figure \ref{fig:bent}).

This type of perturbation and its effects have been
studied in detail in \cite{nikolhosk}. It has been shown that
such an interaction has severe detrimental effects on quantum state transfer in this
network. Numerical results demonstrating the resulting loss of fidelity of quantum state transfer in such a bent chain are presented in  figure \ref{fig:unpr} of section \ref{sec:results}. In our subsequent discussion it will be demonstrated how these detrimental effects can be suppressed efficiently by selective dynamical decoupling.

\section{Basic concepts of dynamical decoupling}\label{sec:methods}
Dynamical decoupling is a method to suppress unwanted interactions between parts of a system. It is based on repeated applications of active control pulses which tend to average out approximately large parts of unwanted interactions in the (assumed to be traceless) Hamiltonian $\Hc$. In an idealized first approximation these control pulses can typically be described by instantaneously applied unitary transformations (bang-bang control \cite{vl98}) which periodically interrupt the free evolution. 

Let the sequence of the unitary operations be denoted by $p_{0},\ldots,p_{m}$ and let us assume that the times of free evolution between consecutive control pulses are all equal and are denoted by $\Delta t$. Thus, after a time $T=m\Delta t$ the time evolution of the system is described by
the unitary transformation
\begin{equation}
U\left(T\right)=p_{m}e^{-i\Hc\Delta t}p_{m-1}e^{-i\Hc\Delta t}\ldots p_{1}e^{-i\Hc\Delta t}p_{0}.
\end{equation}
Introducing the operators
\begin{align}
g_{k} &= p_{k}\cdot p_{k-1}\cdot\ldots\cdot p_{0} \notag \\
\Leftrightarrow p_k &= g_k g_{k-1}^\dag,
\end{align}
this time evolution can be rewritten in the form
\begin{align}
U\left(T\right) & =g_{m}\left(g_{m-1}^{\dagger}e^{-i\mathcal{H}\Delta t}g_{m-1}\right)\ldots\left(g_{0}^{\dagger}e^{-i\mathcal{H} \Delta t}g_{0}\right)\\
 & =g_{m}e^{-i\left(g_{m-1}^{\dagger}\mathcal{H}g_{m-1}\right)\Delta t}\ldots e^{-i\left(g_{0}^{\dagger}\mathcal{H}g_{0}\right)\Delta t} \label{eq:doexponentu} \\
 & \equiv g_m e^{-i \Havg T}. \label{eq:Havg}
\end{align}
In \eref{eq:doexponentu} we used the fact that the transformations $g_{k}$
are unitary and can therefore be moved into the exponent. In \eref{eq:Havg} we identify this time evolution with the one caused by an average Hamiltonian $\Havg$ after the same total time $T$. For the following analysis it is convenient to have the remaining operator $g_m$ equal to the identity, which is, in theory, achievable by choosing the final pulse $p_m = g_{m-1}^\dag$. In cases where this final pulse may be difficult to implement in practice, we can conduct the analysis in the toggled frame induced by the operator $g_m$, where the time evolution is described by the operator $\tilde{U}(T) = g_m^\dag U(T)$. 

Using a Magnus expansion \cite{magnus54}, the average Hamiltonian $\Havg$ can be expanded into a series of terms of increasing order of $\Delta t$, i.e.
\begin{equation}
\overline{\mathcal{H}}=\overline{\mathcal{H}}^{\left(0\right)}+ \overline{\mathcal{H}}^{\left(1\right)}+\ldots \label{eq:Magnus} .
\end{equation}
The lowest order of the Magnus expansion is given by
\begin{equation}
\Havg^{(0)}=\frac{1}{m}\sum_{i=0}^{m-1}g_{i}^{\dagger}\Hc g_{i}, \label{eq:lowest_order}
\end{equation}
and the higher orders depend on $\Delta t$ according to $\Havg^{(k)} = O\left((m \Delta t)^k\right)$.

A set of operators $\{g_i\}_{i=0}^{m-1}$ is traditionally called a decoupling scheme if, to lowest order of $\Delta t$, it eliminates the Hamiltonian $\Hc$, i.e., $\Havg^{(0)}=0$. However, in our case we only want to eliminate parts of $\Hc$ in order to approximately turn the average Hamiltonian $\Havg$ into the ideal Hamiltonian $\Hid$. This is expressed by the decoupling condition
\begin{equation}
\Havg^{(0)}=\frac{1}{m}\sum_{i=0}^{m-1}g_{i}^{\dagger}\mathcal{H}g_{i}=\frac{1}{D}\Hid.\label{eq:podminka}
\end{equation}
We allow for a scaling factor $D$. If $D\neq 1$, its effect can be compensated by rescaling the overall interaction time by the factor $D$. Any set of operators $\{g_i\}_{i=0}^{m-1}$ which fulfils the decoupling condition \eref{eq:podminka} is called a selective dynamical decoupling scheme. 

We emphasize that decoupling is only an approximate method, and that \eref{eq:podminka} guarantees only a
decoupling in first order of $\Delta t$. Several strategies exist to decrease the influence of the higher orders. The simplest strategy is called periodic dynamical decoupling (PDD) and works by repeating the operators of the decoupling scheme. This creates a new decoupling scheme with $m'=2m$ operators of the form
\begin{equation}
g_j' = g_{j (\bmod m)} .
\end{equation}
Assuming that the total interaction time $T'=m' \Delta t'$ remains the same, $T'=T$, then the distance between two pulses is reduced by half, $\Delta t' = \frac{1}{2} \Delta t$. For the time evolution of this new scheme, we find
\begin{align}
U'(T) &= \prod_{i=0}^{m'-1} e^{-i \left(g_i'^\dag \mathcal{H} g_i'\right) \Delta t'}  \notag \\
&= \left( \prod_{i=0}^{m-1} e^{-i \left(g_i^\dag \mathcal{H} g_i\right) \Delta t'}  \right) \left( \prod_{i=0}^{m-1} e^{-i \left(g_i^\dag \mathcal{H} g_i\right) \Delta t'}  \right) \notag \\
&= U(m \Delta t')^2 \equiv \left( e^{-i \Havg T/2} \right)^2 = e^{-i \Havg T} . 
\end{align}
Here, the products are meant to be ordered right to left to match \eref{eq:doexponentu}.
The resulting average Hamiltonian $\Havg$ is formally identical to that of the original scheme, but depends on the reduced $\Delta t'$. This means that the lowest order $\Havg^{(0)}$ remains the same, but the higher orders are smaller due to their scaling with $(\Delta t')^k$.

A more sophisticated strategy is called symmetric dynamical decoupling (SDD). For this strategy, we construct a scheme with $m'=2m$ operators from the original scheme by appending its own reverse. The operators of this scheme are then given by
\begin{equation}
g'_{i}=\begin{cases}
g_{i} & \mathrm{for\,}i<m,\\
g_{2m-1-i} & \mathrm{for\,}i\geq m.
\end{cases}
\end{equation}
This symmetrized decoupling scheme eliminates all the odd orders $\Havg^{(k)}, k \in \{1,3,5,\dots\}$ in the Magnus expansion
\eref{eq:Magnus} \cite{bur81}.
Typically, this improves the scheme's performance significantly. It can then be used with the PDD strategy to reduce the influence of the higher orders even further.

\section{Selective dynamical decoupling schemes for state transfer}
\label{sec:solution}

In order to develop a decoupling scheme to suppress the effects of the unwanted coupling, let us introduce operators 
\begin{equation}
h_{i,j} = \sigma_i^x \sigma_{j}^x + \sigma_i^y \sigma_{j}^y ,
\end{equation}
which allow us to rewrite the Hamiltonians as
\begin{align}
\Hid &= \sum_i B_i \sigma_i^z - \sum_{i} J_{i} h_{i,i+1} , \nonumber \\
\Hc &= \Hid + \gamma h_{\alpha-1,\alpha+1}
\end{align}
Choosing an operator $g = \sigma_i^z$ or $g = \sigma_{j}^z$, a straight-forward calculation yields
\begin{equation}
g^\dag h_{i,j} g = - h_{i,j},
\end{equation}
so that $g$ acts as a time-reversal operator for the Hamiltonian component $h_{i,j}$. If we consider two decoupling scheme operators $g_0 = \id$ and $g_1 = g = \sigma_{\alpha-1}^z$ and insert them into \eref{eq:lowest_order}, 
we find
\begin{align}
\Havgo &= \frac{1}{2} (\Hid + g^\dag \Hid g + \gamma h_{\alpha-1,\alpha+1} -\gamma h_{\alpha-1,\alpha+1}) \\
&= \frac{1}{2} (\Hid + g^\dag \Hid g).
\end{align}
Thus, to lowest order the unwanted coupling appearing in the Hamiltonian \eref{eq:hamiltonian}) is eliminated, independent of the actual strength $\gamma$ of the error. The possibility of schemes removing certain unwanted terms from a Hamiltonian even when they are not exactly known was already established in \cite{fab14,bbcak10}. Unfortunately, the remaining term in $\Havgo$ is not equal to $\Hid$ because $g^\dag \Hid g \neq \Hid$. But we can build on this observation and expand the sequence of operators to get closer to our ultimate goal of simulating $\Hid$.

\subsection{Partial selective dynamical decoupling} \label{ssec:simple}
Let us first of all investigate a selective decoupling scheme which achieves this goal partly.
For this purpose we split the operator $g$ into two operators. One of them acts on the qubit $\alpha-1$ and the other one on qubit $\alpha+1$. Thus, they reverse the sign of $h_{\alpha-1,\alpha+1}$ in a way that the result of \eref{eq:lowest_order} is proportional to $\Hid$. Specifically we choose decoupling operators of the form
\begin{align}
g_0 &= \id, \notag \\
g_1 &= \id, \notag \\
g_2 &= \sigma_1^z  \sigma_3^z \dots \sigma^z_{\alpha-3} \sigma^z_{\alpha-1}, \notag \\
g_3 &= \sigma_{\alpha+1}^z \sigma_{\alpha+3}^z \dots \sigma_{N-2}^z \sigma_N^z,
\end{align}
where $g_2$ acts on qubit $\alpha-1$ and every second qubit before it. Similarly, $g_3$ acts on qubit $\alpha+1$ and every second qubit after it. The unitary transformation $g_2$ induces a time reversal affecting all operators $h_{i,i+1}$ for $i<\alpha$ and $h_{\alpha-1,\alpha+1}$. Analogously, $g_3$ acts as a time reversal operation on all operators $h_{i,i+1}$ with $i\ge \alpha$ and on $h_{\alpha-1,\alpha+1}$. Since two operators  reverse the sign of $h_{\alpha-1,\alpha+1}$, we also need to include $\id$ twice in our decoupling scheme to bring its total sum to zero. Calculating $\Havgo$ for this decoupling scheme yields
\begin{align}
\Havgo &= \sum_i B_i \sigma_i^z - \frac{1}{2} \sum_{i} J_i (\sigma_i^x \sigma_{i+1}^x + \sigma_i^y \sigma_{i+1}^y) .
\end{align}
Compared to $\Hid$, we have all the two-qubit interactions scaled by a factor $\frac{1}{2}$, meaning $D=2$ in \eref{eq:podminka}, which can be accounted for by increasing the interaction time for the transfer by a factor of 2. However, the eigenenergies of the qubits $B_i \sigma_i^z$ are not scaled, so our decoupling scheme is not able to achieve $\Hid$ to lowest order perfectly. If all the $B_i$ are the same, as required for successful state transfer, then the effect of this discrepancy in the scaling is just a relative phase $e^{2iBT}$ which is picked up by the transferred state and could be corrected after the transfer occurred. As such, the discrepancy may be perfectly acceptable in practice, depending on the specifics of the studied system.
However, for cases in which the resulting phase shift cannot be compensated by an appropriate unitary transformation at the end of the quantum state transfer protocol, a complete selective dynamical decoupling scheme is needed which scales both parts of $\Hid$ by the same factor of $\frac{1}{2}$ in the lowest order of the Magnus expansion. In the subsequent section such a selective decoupling scheme is
developed.

\subsection{Complete selective dynamical decoupling}\label{ssec:full}
For a systematic derivation of a complete dynamical decoupling scheme which yields the same scaling for all the parts of $\Hid$, one can use the general method
developed recently by some of the authors in \cite{fab14}. This method is based on similar ideas as involved in the derivation of the partial selective decoupling scheme and generalizes them by a formal procedure that allows to find selective dynamical decoupling schemes by solving an appropriate linear and inhomogeneous system of equations. Instead of recapitulating this general procedure presented in detail in \cite{fab14}, in the following
we present the resulting complete selective dynamical decoupling scheme, discuss its main features, and demonstrate that
it works as intended. In analogy to the previously discussed partial selective dynamical decoupling scheme, the complete scheme involves four unitary operators of the form
\begin{align}
g_0 &= \id \notag \\
g_1 &= \sigma_1^x \sigma_2^x \dots \sigma_{\alpha-2}^x \sigma_{\alpha-1}^x \notag \\
g_2 &= \sigma_{\alpha+1}^y \sigma_{\alpha+2}^y \dots \sigma_{N-1}^y \sigma_N^y \notag \\
g_3 &= \sigma_2^z \sigma_4^z \dots \sigma_{\alpha-3}^z \sigma_{\alpha-1}^z \sigma_\alpha^x \sigma_{\alpha+2}^z \sigma_{\alpha+4}^z \dots \sigma_{N-2}^z \sigma_N^z .
\end{align}
In the unitary operation $g_1$ a $\sigma^x$ operator acts on all qubits up to $\alpha-1$
and in $g_2$ a $\sigma^y$ operator is applied to all qubits starting from $\alpha+1$. 
Since $\sigma^x \sigma^z \sigma^x = \sigma^y \sigma^z \sigma^y = -\sigma^z$ both the $\sigma^x$ and the $\sigma^y$ operators act as time reversal operators for the eigenenergy terms of the qubits. Analogously, in the sum of 
\eref{eq:lowest_order} the operators $g_1$ and $g_2$ introduce minus signs in the eigenenergy terms of all affected qubits. The operator $g_3$ involves a single $\sigma^x$ operator acting on qubit $\alpha$ at the bend. Therefore, for each qubit $i$ there is a decoupling operator yielding a minus sign in the term $B_i \sigma_i^z$ of the sum of \eref{eq:lowest_order} and in addition there are three operators for each qubit yielding a positive sign. Thus, all of the terms $B_i \sigma_i^z$ are weakened by a scaling factor $D=2$ as needed. 

We still need to confirm that the unwanted coupling between qubits $\alpha-1$ and $\alpha+1$ is removed and the remaining two-qubit couplings are scaled by a factor of $D=2$. Let us first ignore the qubit $\alpha$ at the bend and let us focus on the rest of the qubit chain. In view of the relation 
\begin{equation} \label{eq:effects_xx}
\sigma_i^x \sigma_{i+1}^x h_{i,i+1} \sigma_i^x \sigma_{i+1}^x = \sigma_i^y \sigma_{i+1}^y h_{i,i+1} \sigma_i^y \sigma_{i+1}^y = h_{i,i+1},
\end{equation}
the operators $g_1$ and $g_2$ yield positive signs in the couplings $h{i,i+1}$ for 
$i \in [1,\alpha-2] \cup [\alpha+1, N-1]$. The operator $g_3$, however, yields a negative sign in these couplings. Therefore, we obtain a scaling of these couplings with $D=2$ as expected. 
The relevant couplings at the bend are $h_{\alpha-1,\alpha}$, $h_{\alpha,\alpha+1}$ and $h_{\alpha-1,\alpha+1}$ the latter of which we want to remove to lowest order of the Magnus expansion. 
The operators $g_j$ transform these couplings in the following way
\begin{align}
g_1 h_{\alpha-1,\alpha} g_1 &= \sigma_{\alpha-1}^x \sigma_{\alpha}^x - \sigma_{\alpha-1}^y \sigma_{\alpha}^y, \notag \\ 
g_2 h_{\alpha-1,\alpha} g_1 &= \sigma_{\alpha-1}^x \sigma_{\alpha}^x + \sigma_{\alpha-1}^y \sigma_{\alpha}^y, \notag \\ 
g_3 h_{\alpha-1,\alpha} g_3 &= -\sigma_{\alpha-1}^x \sigma_{\alpha}^x + \sigma_{\alpha-1}^y \sigma_{\alpha}^y, \notag \\ 
g_1 h_{\alpha,\alpha+1} g_1 &= \sigma_{\alpha}^x \sigma_{\alpha+1}^x + \sigma_{\alpha}^y \sigma_{\alpha+1}^y, \notag \\ 
g_2 h_{\alpha,\alpha+1} g_2 &= -\sigma_{\alpha}^x \sigma_{\alpha+1}^x + \sigma_{\alpha}^y \sigma_{\alpha+1}^y, \notag \\ 
g_3 h_{\alpha,\alpha+1} g_3 &= \sigma_{\alpha}^x \sigma_{\alpha+1}^x - \sigma_{\alpha}^y \sigma_{\alpha+1}^y, \notag \\ 
g_1 h_{\alpha-1,\alpha+1} g_1 &= \sigma_{\alpha-1}^x \sigma_{\alpha+1}^x - \sigma_{\alpha-1}^y \sigma_{\alpha+1}^y, \notag \\ 
g_2 h_{\alpha-1,\alpha+1} g_2 &= -\sigma_{\alpha-1}^x \sigma_{\alpha+1}^x + \sigma_{\alpha-1}^y \sigma_{\alpha+1}^y, \notag \\ 
g_3 h_{\alpha-1,\alpha+1} g_3 &= -\sigma_{\alpha-1}^x \sigma_{\alpha+1}^x - \sigma_{\alpha-1}^y \sigma_{\alpha+1}^y. 
\end{align}
Using these results and looking at the sum of \eref{eq:lowest_order} we notice
that the coupling $h_{\alpha-1,\alpha+1}$ is indeed eliminated 
as the applications of operators $g_0$ and $g_3$ cancel each other.
Similarly, this is valid for the operators $g_1$ and $g_2$. In the case of the other two couplings, i.e. $h_{\alpha-1,\alpha}$ and $h_{\alpha,\alpha+1}$, the couplings remain in the result of the sum with a factor of ${1}/{2}$ each as required. Therefore, the new scheme fulfils the necessary selective dynamical decoupling condition \eref{eq:podminka} with a scaling factor of $D=2$.

\subsection{A practical decoupling scheme} \label{ssec:practical}
The two decoupling schemes presented so far share a common drawback. They both require $\sigma^z$ pulses, which are typically hard to implement experimentally. It would therefore be beneficial to have a decoupling scheme that employs only $\sigma^x$ and $\sigma^y$ pulses, both of which are usually much easier to implement. We were able to find such a scheme, which consists of the following four operators:
\begin{align}
g_0 &= \id, \notag \\
g_1 &= \sigma_1^x \sigma_2^x \dots \sigma_{\alpha}^x \sigma_{\alpha+1}^y \sigma_{\alpha+2}^x \sigma_{\alpha+3}^y \dots \sigma_N^x , \notag \\
g_2 &= \id, \notag \\
g_3 &= \sigma_1^y \sigma_2^x \sigma_3^y \dots \sigma_{\alpha-2}^x \sigma_{\alpha-1}^y \sigma_{\alpha}^x \sigma_{\alpha+1}^x \dots \sigma_N^x .
\end{align}
The operator $g_1$ applies $\sigma^x$ to all qubits up to the bend position $\alpha$, then alternates between $\sigma^y$ and $\sigma^x$ for the remaining qubits. The operator $g_3$ is basically a mirror of $g_1$ and applies $\sigma_x$ to all qubits starting from the bend position $\alpha$ to the end of the chain, but alternates between $\sigma^y$ and $\sigma^x$ before the bend. Both operators act on all qubits at the same time. In practical realizations, only the pulse phase would need to be altered for the individual qubits to differentiate between $\sigma^x$ and $\sigma^y$ pulses, whereas the source of the pulses may be shared by all qubits, allowing for potentially easier implementation.

With the result from \eref{eq:effects_xx} and the additional relations 
\begin{align}
& \sigma_i^x \sigma_{i+1}^y h_{i,i+1} \sigma_i^y \sigma_{i+1}^x = \sigma_i^y \sigma_{i+1}^x h_{i,i+1} \sigma_i^x \sigma_{i+1}^y = -h_{i,i+1}, \notag \\
& \sigma_i^x \sigma_i^z \sigma_i^x = \sigma_i^y \sigma_i^z \sigma_i^y = -\sigma_i^z
\end{align}
we can easily verify that the lowest order of the average Hamiltonian takes the form
\begin{equation}
\Havgo = - \frac{1}{2} \sum_{i} J_i (\sigma_i^x \sigma_{i+1}^x + \sigma_i^y \sigma_{i+1}^y) .
\end{equation}
Just like with the previous two schemes, the interactions between the qubits are preserved with a scaling of $D=2$, while the additional coupling at the bend is eliminated in the lowest order. However, this scheme also eliminates the eigenenergy terms $B_i \sigma_i^z$ to lowest order. This means that the relative phase shift from the transfer depends entirely on the length of the chain $N$ and is given by $e^{i\varphi} = (-i)^{N-1}$. 

Since $\Havgo$ does not depend on specific values of the $B_i$, the eigenenergies are eliminated even if they are not uniform. This offers a practical advantage over the other two schemes: since non-uniform eigenenergies disturb the transfer, this scheme is robust against this kind of disorder and allows the state transfer to complete successfully even in the presence of diagonal disorder. The occuring phase shift is predictable and therefore easily corrected after the transfer. For this reason and for the lack of $\sigma^z$ decoupling operators, we believe this scheme to be the best suited for a practical implementation.

\section{Numerical simulations} \label{sec:results}
To investigate the effectiveness of the proposed selective dynamical decoupling schemes we present numerical simulations of the dynamics of linear qubit chains governed by the Hamiltonian of \eref{eq:hamiltonian} with the coupling strengths of \eref{eq:couplings}. For the additional coupling strength $\gamma$ we chose
\begin{equation}
\gamma = 0.4\max\left\{ \frac{\lambda}{2}\sqrt{\left(\alpha-1\right)\left(N-\alpha+1\right)}, \frac{\lambda}{2}\sqrt{\left(\alpha\right)\left(N-\alpha\right)}\right\} ,\label{eq:gamma}
\end{equation}
and $\alpha$ to be at or close to the middle of the chain. This choice is sufficient to cover most of the interesting situations \cite{nikolhosk}.

Since the Hamiltonian and the decoupling procedure preserve the total number of excitations to quantify the quality of the state transfer it is sufficient to consider a single excitation transfer \cite{broughnikoljex}. In this scenario the first qubit of the network is prepared in its excited state and the rest is in the ground state, i.e.
\begin{equation}
	\ket{\psi_i} = \ket{1}\ket{0}\ldots\ket{0}.
\end{equation}
Perfect quantum state transfer occurs if there exists a time $T$ after which the system evolves to
the state
\begin{equation}
	\ket{\psi(T)}=\ket{\psi_f} = \ket{0}\ldots\ket{0}\ket{1}.
\end{equation}
If we considered a general linear combination to be present at the first qubit, we would only get a relative phase for all three schemes at the end, which can be found in the respective subsections of Section \ref{sec:solution} calculated explicitly. Let us note here that we performed numerical simulations of the phase change for all considered schemes and the given formulas give the right phase change up to a very good order.

We measure the transfer quality by means of the state fidelity $F$, which in our case is given by
\begin{equation}
	F(t) = \abs{\braket{\psi_f}{\psi(t)}}, \quad F(t) \in [0,1].
\end{equation}
Perfect state transfer has occurred after time $T$ if $F(T)=1$.

\begin{figure}[ht]
\includegraphics[width=7.5cm]{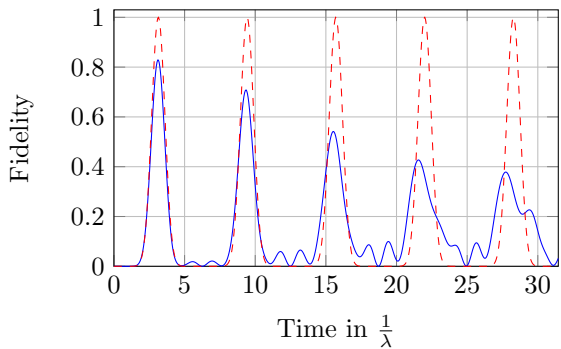}
\caption{Fidelity $F$ as a function of time (in units of $1/\lambda$) for
a 10-qubit chain with Hamiltonian \eref{eq:hamiltonian}: \\
Ideal unperturbed qubit chain with $\gamma = 0$ (red dashed line); \\
Perturbed bent qubit chain with $\gamma$ of \eref{eq:gamma} and $\alpha=5$ (blue line).}
\label{fig:unpr}
\end{figure}

Let us first of all look at the time evolution resulting from the Hamiltonian 
\eref{eq:hamiltonian} without any selective dynamical decoupling 
applied. For a chain of $10$ qubits, numerical results are depicted in figure 
\ref{fig:unpr}. As expected, the fidelity never reaches the optimal value of unity. 
After the time $T=\pi/\lambda$ the fidelity of the bent $10$-qubit chain assumes its maximum at $\approx 0.83$. This is the time where we expect perfect quantum state transfer to happen under ideal conditions.

\subsection{Complete selective dynamical decoupling scheme}
Let us now investigate how
well the complete solution
presented in section \ref{ssec:full} protects quantum state transfer in a bent linear qubit chain.
For this purpose, we will simulate the time evolution resulting from applying the scheme for a specific number of repetitions during the transfer time $T$.

\begin{figure}[ht]
\includegraphics[width=7.5cm]{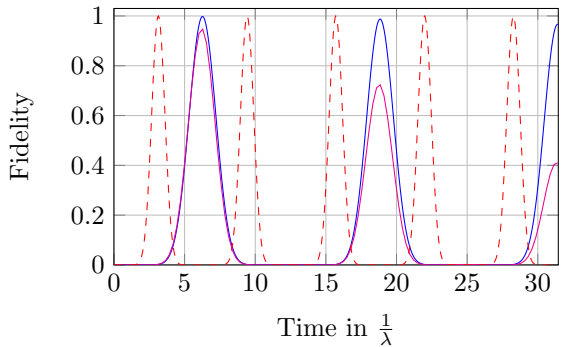}
\caption{Fidelity $F$ as a function of time (in units of $1/\lambda$) for
a 10-qubit bent chain with Hamiltonian \eref{eq:hamiltonian}
protected by the complete selective dynamical decoupling scheme: 
12 repetitions of the complete scheme with a total number of 48 pulses per $\pi/\lambda$ (magenta lower line);
60 repetitions of the complete scheme with a total number of 240 pulses per $\pi/\lambda$ (blue line). The red dashed line shows the time evolution of an ideal unperturbed 10-qubit chain.}
\label{fig:10q240}
\end{figure}

Figure \ref{fig:10q240} presents numerical results obtained for a bent 10-qubit chain under the protecting influence of the complete dynamical decoupling scheme. Two cases are depicted with different frequencies of the applied control operations.  In the first case (magenta lower line) the selective decoupling scheme is repeated 12 times with a total number of 48 pulses per $\pi/\lambda$ period required for perfect state transfer in the ideal unperturbed case.
In the second case the selective dynamical decoupling scheme is repeated 60 times with a total number of 240 pulses per $\pi/\lambda$ period. We notice that now the fidelity peak occurs after a time $2\pi/\lambda$ which originates from the decoupling scheme's time scaling factor of $D=2$. It is also apparent that in both cases the fidelity maximum is higher than in the unprotected case; for 12 repetitions it reaches a value of  $F\approx 0.947$ and for 60 repetitions it reaches $F\approx 0.998$. The beneficial influence of higher repetitions is particularly apparent at the subsequent fidelity peaks. In the case of 60 repetitions the achievable fidelities at these maxima are still close to unity. However, for practical purposes the first fidelity maximum at time
$t = 2\pi/\lambda$ is the most relevant one. In actual experiments the achievable number of control pulses is likely to be limited, so it is important to find a reasonable balance between the required number of control pulses and the achieved transfer fidelity.

\begin{figure}[ht]
\includegraphics[width=7.5cm]{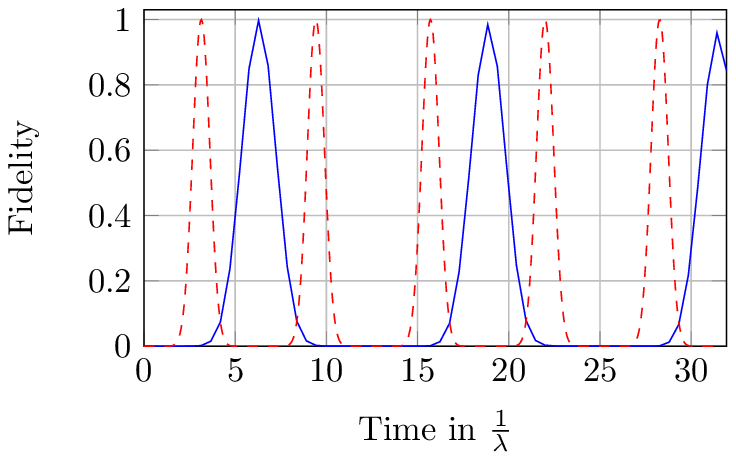}
\caption{Fidelity $F$ as a function of time (in units of $1/\lambda$) for
a 10-qubit bent chain with Hamiltonian \eref{eq:hamiltonian}
protected by the symmetrized complete selective dynamical decoupling scheme with 
6 repetitions of the symmetrized complete scheme involving a total number of 48 pulses per $\pi/\lambda$ (blue line); the red dashed line shows the time evolution of an ideal unperturbed 10-qubit chain.}
\label{fig:10q48sym}
\end{figure}

To improve the performance of the scheme, we can employ the SDD decoupling strategy. Even though the symmetrized scheme consists of twice as many operations, it should require fewer repetitions than the original sequence to achieve a given degree of error suppression.
Numerical results are presented in figure \ref{fig:10q48sym} for the symmetrized selective dynamical decoupling scheme. Thereby 6 repetitions of the symmetric scheme have been performed involving a total number of 48 control pulses per $\pi/\lambda$ period. This is the same number of control pulses as used for obtaining the black curve of figure \ref{fig:10q240}. The fidelity maximum is now closer to unity at a value of $F\approx 0.997$ which is comparable to simulations involving the original selective dynamical decoupling scheme with the significantly larger number of 240 control pulses. This demonstrates that the symmetrized version of the complete selective dynamical decoupling scheme performs significantly better.

\subsection{Partial selective dynamical decoupling scheme}
Let us now investigate the performance of the practical selective dynamical decoupling scheme introduced in section \ref{ssec:simple}. It is also suitable for protecting quantum state transfer on a bent qubit chain with the caveat that the qubits' eigenenergies are not properly rescaled. In general this leads to a relative phase change during a quantum state transfer which has to be taken into account. Whether or not this is a problem in practical applications depends on experimental circumstances. In the following it will be demonstrated that in some respects this simpler partial selective dynamical decoupling scheme performs even better than the symmetrized complete scheme. This feature is attractive for practical application provided the resulting phase change can be corrected at the end of a quantum state transfer by other means. 
Note that the phase change is only relevant if transferring a superposition state $\alpha \ket{0} + \beta \ket{1}$. The transfer fidelity for the state $\ket{1}$, which we use in our simulations, is unaffected.

\begin{figure}[ht]
\includegraphics[width=7.5cm]{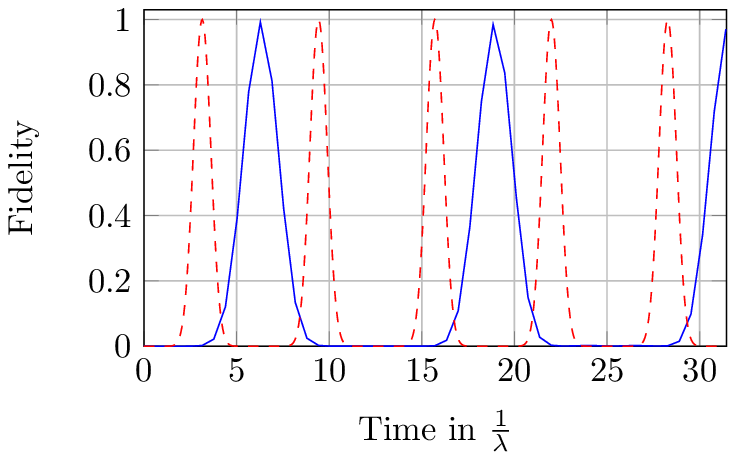}
\caption{Fidelity $F$ as a function of time (in units of $1/\lambda$) for
a 10-qubit bent chain with Hamiltonian \eref{eq:hamiltonian}
protected by the partial selective dynamical decoupling scheme with 
5 repetitions of the partial scheme involving a total number of 20 pulses per $\pi/\lambda$ (blue line); the red dashed line shows the time evolution of an ideal unperturbed 10-qubit chain.}
\label{fig:10q20simple}
\end{figure}

Figure \ref{fig:10q20simple} shows the influence of this partial selective dynamical decoupling scheme on the dynamics of a bent 10-qubit chain. In this example 5 repetitions of this scheme are used which involve a total number of 20 control pulses per $\pi/\lambda$ period. This is less than half the number of control pulses used in the symmetric case depicted in figure \ref{fig:10q48sym}. Yet the performance is quite comparable. The fidelity maximum reaches a value of $F\approx 0.992$.

\subsection{Practical decoupling scheme}
\begin{figure}[ht]
\includegraphics[width=7.5cm]{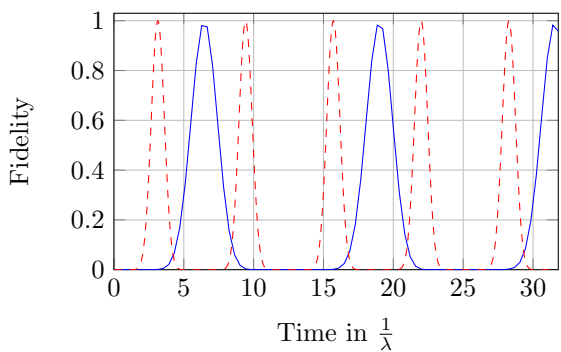}
\caption{Fidelity $F$ as a function of time (in units of $1/\lambda$) for
a 10-qubit bent chain with Hamiltonian \eref{eq:hamiltonian}
protected by the practical dynamical decoupling scheme with 
8 repetitions of the practical scheme involving a total number of 32 pulses per $\pi/\lambda$ (blue line); the red dashed line shows the time evolution of an ideal unperturbed 10-qubit chain.}
\label{fig:10q32xy}
\end{figure}

In this subsection we investigate the performance of the practical decoupling scheme from Subsec. \ref{ssec:practical}, which does not make use of the $\sigma^z$ pulses at all. A representative case of the time evolution numerically simulated is plotted in figure \ref{fig:10q32xy}. From our simulations it seems that the number of pulses needed for quantitatively similar effects as the previous two schemes lies somewhere in between the two other schemes, somewhat closer to the number of pulses needed with the complete scheme. 

\subsection{Selective dynamical decoupling in long qubit chains}
In practical realizations of selective dynamical decoupling schemes 
the number of control pulses that can be implemented may be limited. In the following we investigate how the minimal number of control pulses necessary for achieving a satisfactory transfer fidelity scales with the number of qubits in bent qubit chains. For this purpose
we concentrate on an achievable transfer fidelity of $F=0.95$ at the first maximum of the quantum state transfer protocol in qubit chains involving up to eleven qubits and
determine the minimal number of pulses required to reach this transfer fidelity. 

\begin{figure}[ht]
\includegraphics[width=7.5cm]{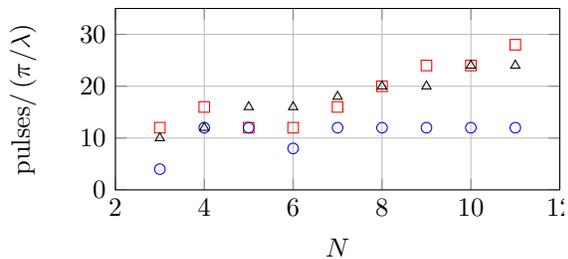}
\caption{Minimum number of control pulses per $\pi/\lambda$ period to achieve a transfer fidelity 
$F\ge 0.95$ and its dependence on the number $N$ of qubits in a bent chain with
$\gamma$ given by \eref{eq:gamma} and with $\alpha$ positioned in the middle of the chain: 
Symmetrized complete selective dynamical decoupling scheme (red squares); partial selective dynamical decoupling scheme (blue circles); practical decoupling scheme (black triangles)}
\label{fig:kvsn}
\end{figure}

Numerical results are depicted in figure \ref{fig:kvsn} for all the partial selective dynamical decoupling scheme, practical decoupling scheme and the symmetrized complete scheme. Apart from small qubit chains the number of control pulses required in the symmetrized complete scheme and the practical scheme grow approximately linearly with the number $N$ of qubits of the chain. For $N<6$ somewhat more control pulses are required which may originate from the disturbance being too close to the ends of the chain and thus having a particularly strong impact. The partial selective dynamical decoupling scheme also exhibits this phenomenon. But for longer qubit chains it requires an approximately constant number of 12 control pulses per $\pi/\lambda$ period. We expect, however, that for even larger qubit chains the number of required control pulses will also eventually grow linearly with $N$, albeit possibly with a smaller slope than the symmetrized complete selective decoupling scheme.

In \cite{nikolhosk} it has been demonstrated that the effect of the perturbing additional coupling at the bend of a linear qubit chain diminishes with increasing numbers $N$ of qubits of the chain. In view of the linearly increasing number of pulses necessary to counteract the influence of the disturbance we expect that for very long qubit chains the effort required to successfully implement decoupling may no longer be worth the expected benefits. Therefore, the presented selective dynamical decoupling schemes are expected to be particularly valuable for protecting quantum state transfer in quantum networks of intermediate sizes which are of interest for current realisations of quantum registers.

\subsection{Imperfect pulses}
Since we consider the additional coupling in the network to be a result of some imperfection or defect, it is important to investigate how the suggested schemes work under imperfect conditions themselves. So far, we have assumed that both the pulses and the timing between pulses are perfect. In this section, we will study the effects of two different sources of errors. The first are imperfections in the timing of the pulses, which we model by replacing the constant $\Delta t$ with random values from a Gaussian distribution with mean value $\mu = \Delta t$ and standard deviation $\sigma = q \Delta t$. The second are systematic errors in the applied Pauli pulses where we replace the perfect pulses $\sigma^i$ with an imperfect pulse $\sigma^i e^{-i \theta \sigma^i}$. Here, $\theta$ can be seen as a rotational offset  when viewing the effects of the Pauli operators on the Bloch sphere. A value of $\theta=0$ corresponds to the ideal pulse.

\begin{figure}[ht]
\subfigure[First three peaks in Fidelity]{
	\includegraphics[width=7.5cm]{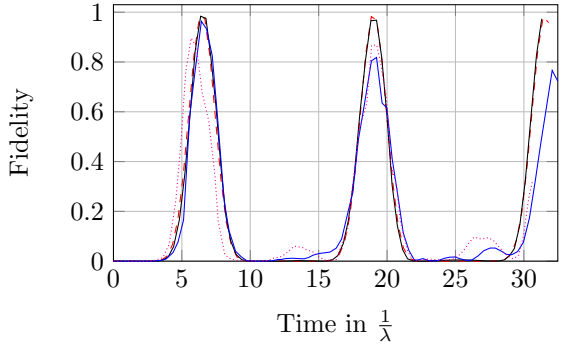}
}\\ 
\subfigure[Cut out of the first fidelity peak]{
	\includegraphics[width=7.5cm]{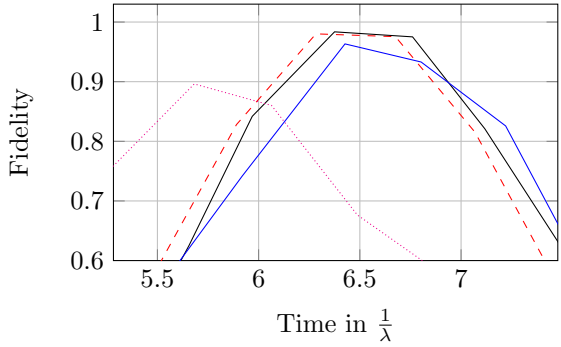}
}
\caption{Fidelity $F$ as a function of time (in units of $1/\lambda$) for
a 10-qubit bent chain with Hamiltonian \eref{eq:hamiltonian}
protected by the practical scheme with 
8 repetitions of the scheme involving a total number of $\approx 32$ pulses per $\pi/\lambda$ seconds with pulses placed imperfectly, randomly in time. Notice the imperfect timings and drops in the first fidelity peak depending on the standard deviation $\sigma$:\\
 red dashed line: $\sigma = 0$, exactly placed, perfect pulses\\
 black solid line: $\sigma = 0.1 \Delta t$ very similar result to perfectly placed pulses\\
 blue solid line: $\sigma = 0.3 \Delta t$, significant drop in all fidelity peaks behind the first peak\\
 magenta dotted line: $\sigma = 0.5 \Delta t$, biggest drop in the first fidelity peak}
 \label{fig:random}
\end{figure}

The results for imperfect timings can be seen for the practical scheme in figure \ref{fig:random}. We have also run simulations for the other schemes, and the results are similar in the sense that the fidelity peaks begin to drop significantly once $q \geq 0.2$ and do not change very much for $q \in \left[ 0, 0.2\right)$. In other words: if 95.4\% of pulses happen between $\Delta t \pm 0.4 \Delta t$ with Gaussian distribution around $\Delta t$, the decoupling schemes generally perform close to the case of perfect timing. 

\begin{figure}[ht]
\subfigure[First three peaks in Fidelity]{
	\includegraphics[width=7.5cm]{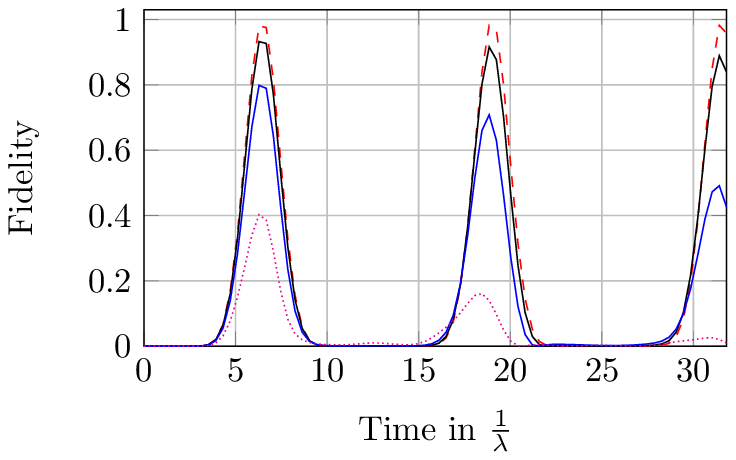}
}\\ 
\subfigure[Cut out of the first fidelity peak]{
	\includegraphics[width=7.5cm]{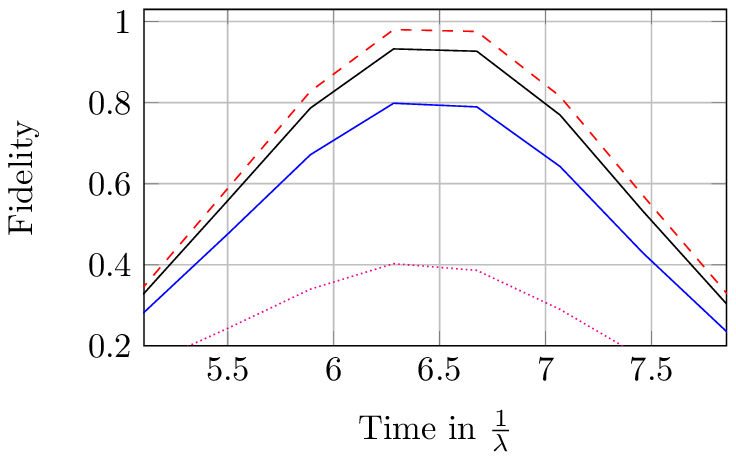}
}
\caption{Fidelity $F$ as a function of time (in units of $1/\lambda$) for
a 10-qubit bent chain with Hamiltonian \eref{eq:hamiltonian}
protected by the practical scheme with 
8 repetitions of the scheme involving a total number of $32$ pulses per $\pi/\lambda$ seconds with imperfect pulses - a systematic error is present with all the pulses. Different $\theta$'s were selected for the four simulations. Notice the drop in the first fidelity peak and the consequent peaks as well:\\
 red dashed line: $\theta = 0$, perfect pulses\\
 black solid line: $\theta = 0.05 \Delta t$, very similar result to perfect pulses\\
 blue solid line: $\theta = 0.1 \Delta t$, significant drop in all fidelity peaks\\
 magenta dotted line: $\theta = 0.2 \Delta t$, fidelity roughly 0.4 even at the first peak}
 \label{fig:theta}
\end{figure}

For the systematic errors, the results of the practical scheme are shown in figure \ref{fig:theta}. Judging from our simulations, all three schemes were more sensitive to this kind of systematic error than to the randomized timings. In order to keep the first fidelity peak above 0.9, $\theta$ should be kept below $0.1 \Delta t$.

Our simulation results show that there is a reasonable margin for error in the implementation of the decoupling schemes. The systematic error proved to be slightly more problematic, which is to be expected, since a statistical error can average itself out to a certain extent over time. 

\section{Conclusions} \label{sec:conclusion}
Three selective dynamical decoupling schemes have been presented which are capable of suppressing unwanted interactions occurring at a bend of a linear qubit chain. Such a scenario occurs naturally if a chain is formed along a two- or three-dimensional grid of qubits, for example. The selective dynamical decoupling schemes presented weaken the overall Hamiltonian strength by a factor of $1/2$ which has to be compensated by increasing the interaction time in order to achieve quantum state transfer. The additional interaction at the bend is strongly suppressed allowing the chain to work ideally as if the bend was not there. The quality of the suppression depends on the frequency of the applied control pulses, with higher frequency implying better error suppression.

Numerical simulations have been presented for state transfer involving qubit chains of varying lengths. They demonstrate the effectiveness of the selective dynamical decoupling schemes presented. We have also investigated the required minimal number of decoupling pulses to achieve a good transfer fidelity. In the case of the simpler decoupling scheme twelve pulses per $\pi/\lambda$ period have already been sufficient for achieving a satisfactory error suppression in qubit chains of up to eleven qubits. However, with increasing length of the qubit chains we expect
a linear increase of the required number of pulses. This dependence makes our selective dynamical decoupling schemes particularly suitable for applications involving short to intermediate-sized qubit chains, which are relevant in current implementations of quantum registers. On longer chains, the cost of the decoupling method, including the implementation of individual controls on each qubit, may well exceed the benefits.

It should be pointed out that even though we have concentrated on a specific quantum state transfer protocol with specific coupling strengths, the developed selective dynamical decoupling schemes do not depend on the particular coupling strengths as described by \eref{eq:couplings}. There are other choices of coupling strengths implementing quantum state transfer which can be protected by the presented selective dynamical decoupling schemes as well. In fact, the schemes presented should also work for Hamiltonians of the form \eref{eq:HHamiltonian} which have been designed for tasks different from quantum state transfer.

\ack
The authors acknowledge  governmental support by RVO 68407700 and GACR 13-33906S, by SGS13/217/OHK4/3T/14, and support by CASEDIII and by the BMBF-project Q.com.

\FloatBarrier 

\end{document}